\documentclass[aps,prl,twocolumn,showpacs,superscriptaddress,10pt]{revtex4-1} 
\usepackage{amsmath,amssymb,graphicx}

\usepackage{dcolumn}
\usepackage{braket}
\usepackage{amssymb}
\usepackage{amsmath}
\usepackage{latexsym}
\usepackage{mathrsfs}
\usepackage{arydshln}
\usepackage[sans]{dsfont}
\usepackage{epsfig}
\usepackage{epstopdf}

\newcommand{\bitem}{\begin{itemize}}
\newcommand{\fitem}{\end{itemize}}
\newcommand{\beq}{\begin{equation}}
\newcommand{\eeq}{\end{equation}}
\newcommand{\beqa}{\begin{eqnarray}}
\newcommand{\eeqa}{\end{eqnarray}}

\begin{document}

\title{Optical pattern formation with a 2-level nonlinearity}

\author{A. Camara, R. Kaiser, G. Labeyrie\footnote{To whom correspondence should be addressed.}}
\affiliation{Institut Non Lin\'{e}aire de Nice, UMR 7335 CNRS, 1361 route des Lucioles, 06560 Valbonne, France}
\author{W.J. Firth, G.-L. Oppo, G.R.M. Robb, A.S. Arnold, T. Ackemann}
\affiliation{SUPA and Department of Physics, University of Strathclyde, Glasgow G4 0NG, Scotland, UK}

\begin{abstract}
We present an experimental and theoretical investigation of spontaneous pattern formation in the transverse section of a single retro-reflected laser beam passing through a cloud of cold Rubidium atoms. In contrast to previously investigated systems, the nonlinearity at work here is that of a 2-level atom, which realizes the paradigmatic situation considered in many theoretical studies of optical pattern formation. In particular, we are able to observe the disappearance of the patterns at high intensity due to the intrinsic saturable character of 2-level atomic transitions.

\end{abstract}

\maketitle

Spontaneous pattern formation from a homogeneous state is a widespread phenomenon in nonlinear systems out of equilibrium~\cite{Cross1993,Meinhardt1992}. Originating from fields such as chemistry~\cite{Turing1952,Ouyang1991} and hydrodynamics~\cite{Rayleigh1916,Benard1900,Cross1993}, the study of pattern formation has known a rapid development in optics starting from the 80's~\cite{Lugiato1999,Arecchi1999}. Paradigmatic examples such as a Kerr medium~\cite{Yariv1977,Firth1990} or a collection of 2-level atoms at rest~\cite{GrynbergTh1988} were considered in early theoretical studies. Various nonlinear systems, either active such as lasers or photorefractive oscillators~\cite{Tamm1988,Arecchi1999,Arecchi1990}, or passive such as liquid crystals~\cite{LiquidCrystals,Neubecker1995}, were used to realize the first experiments. In some range of experimental conditions, these nonlinear materials mimic ideal systems such as the Kerr medium but one usually lacks a complete theoretical description of the light-matter interaction. Hot atomic vapors were also extensively employed to study various optical instabilities~\cite{GrynbergExp1988,Grynberg1994,Ackemann1994,Ackemann2001,Dawes2005}. There, a theoretical description of the light-atom interaction is available, but the specific experimental conditions (Doppler broadening, hyperfine structure, ballistic or diffusive motion of the atoms) considerably complicate the interpretation. Finally, cold atomic samples started recently to be employed in optical pattern formation~\cite{Greenberg2011,Greenberg2012,Labeyrie2014}.

\begin{figure}
\begin{center}
\includegraphics[width=1\columnwidth]{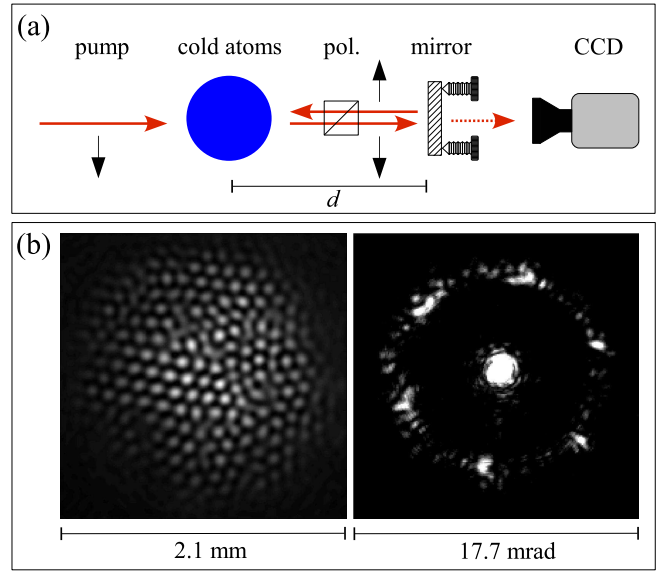}
\caption{(Color on line) Observation of patterns. (a) Experimental scheme. (b) Typical single-shot light distributions observed in the transverse instability regime, in the near (left) and far (right) field. The pump parameters are: $ I = 0.47$ W/cm$^2$ and $\delta = + 6.5~\Gamma$.}
\label{fig1}
\end{center}
\end{figure}

We have identified in our single feedback mirror experiments three distinct mechanisms that lead to the spontaneous formation of patterns. The first one, the optomechanical mechanism, is specific to cold atoms and relies on the spatial bunching of the atoms under the action of dipole forces. This mechanism is very efficient at ultracold temperature and leads to spectacular self-organization phenomena~\cite{Baumann2010}. The presence of a Zeeman structure in the atomic ground state (spin degree of freedom) allows for optical pumping, i.e. a redistribution between populations or creation of coherences between Zeeman substates, in particular within the ground-state. This mechanism is responsible for the polarization instabilities studied in hot atomic vapors~\cite{Gauthier1988,Grynberg1994,Aumann1997,Ackemann2001,Dawes2005}. Finally, under specific experimental conditions the atoms behave as 2-level systems and the optical nonlinearity is only due to the saturation of the electric dipole transition i.e population transfer between ground and excited states. This is the situation studied in this paper, which realizes the paradigmatic theoretical description of a homogeneously-broadened 2-level atomic transition~\cite{GrynbergTh1988}. To our knowledge, the only previous experimental investigation of patterns due to a saturable nonlinearity was achieved in a hot sodium vapor~\cite{GrynbergExp1988}, and did not report the observation of the vanishing of the effect at large saturations. We present in this paper a detailed experimental investigation of this 2-level instability, and obtain a qualitative and quantitative agreement with a theoretical model based only on the microscopic description of the atom-light interaction.

The experimental setup, sketched in Fig.~\ref{fig1}(a), exploits the single-mirror feedback scheme~\cite{Firth1990} (see~\cite{Labeyrie2014} for details). A Gaussian laser beam (referred to as ``the pump'' in the following) of waist $w = 1.47$ mm and wavelength $\lambda = 780.2$ nm is sent through a cold ($T = 200~\mu$K) cloud of Rb$^{87}$ atoms, released from a large magneto-optical trap (MOT). The cloud has a typical size of 9 mm FWHM along the pump propagation axis and contains 10$^{11}$ atoms. The resulting optical density (OD), for a weak beam on resonance with the $F = 2 \rightarrow F^{\prime} = 3$ transition, is around 210. The linearly-polarized pump beam is retro-reflected by a mirror located at a distance $d$ after the cloud (the vertical arrows in Fig.~\ref{fig1}(a) indicate the polarization of the beams). We use an imaging telescope (not represented in Fig.~\ref{fig1}(a)) located between the MOT and the mirror to create a ``virtual mirror'', which provides an access to negative values of $d$~\cite{Ciaramella1993}. The overall reflectivity of the feedback system is around $95\%$ (cloud absorption not included).

By selecting a short (duration $\leq 1~\mu$s) pump pulse, we can neglect the optomechanical nonlinearity which requires tens of $\mu$s since the atoms have to move over distances of the order of $\approx 100 \mu$m~\cite{Labeyrie2014}. Other mechanisms, relying on Zeeman internal degrees of freedom, also lead to a transverse instability and are currently under study in our groups. However, in the setup studied in the present paper a polarizer placed inside the feedback loop (Fig.~\ref{fig1}(a)) guarantees that the feedback only occurs in the polarization channel parallel to the incident pump polarization, and prevents the occurrence of a polarization instability linked to Zeeman degrees of freedom. This is confirmed by the fact that almost no light is detected in the polarization channel orthogonal to the pump polarization. Throughout this work we are thus left with the simplest, 2-level nonlinearity corresponding to the following expression for the refractive index of our cloud of cold atoms:
\begin{equation}
n = 1 - \frac{3\lambda^3}{4\pi^2} \, \frac{\delta/\Gamma}{1+ (2\delta/\Gamma)^2} \, \frac{\rho}{1+s}
\label{index}
\end{equation}
where $\delta = \omega_l -\omega_0$ is the detuning between the laser and atomic frequencies, $\Gamma = 2\pi \times 6.06$ MHz is the natural width and $\rho$ denotes the spatial atomic density. The nonlinearity arises from the presence in this expression of the saturation parameter $s = \frac{I}{I_{sat}} \frac{1}{1+4 (\delta / \Gamma)^2}$, where $I_{sat} = 1.67$ mW/cm$^2$ is the saturation intensity. For $s \ll 1$, the gas exhibits a Kerr-like behavior $n \simeq n_0 + n_2~I$, where $n_0$ is the linear refractive index and $n_2$ the nonlinear one. Importantly, for $s \gg 1$ the nonlinearity vanishes and the instability is expected to disappear. In the Kerr regime, the nonlinearity is ``self-focusing'' ($n_2 > 0$) for $\delta > 0$ and ``self-defocusing'' for $\delta < 0$. Note that Eq.~\ref{index} only describes the real part of the complex refractive index, which is responsible for the instability. However, (nonlinear) absorption is also present and included in our theoretical analysis.

\begin{figure}
\begin{center}
\resizebox{1.0\columnwidth}{!}{\includegraphics{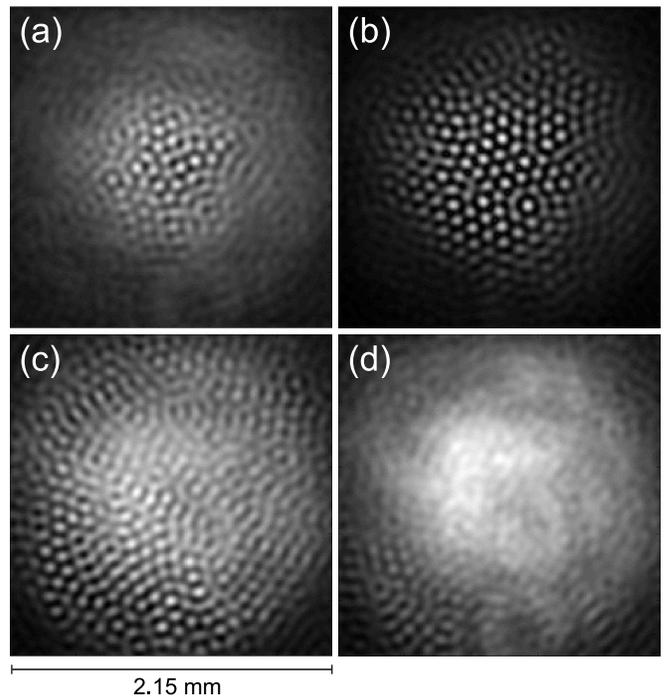}} 
\caption{Saturation of the instability. We show the evolution of the patterns as the pump intensity is increased : a) $ I = 0.24$ W/cm$^2$, (b) $ I = 0.47$ W/cm$^2$, (c) $ I = 1.41$ W/cm$^2$ and (d) $ I = 4.24$ W/cm$^2$. The detuning is $\delta = + 6.5~\Gamma$.}
\label{fig2}
\end{center}
\end{figure}

Fig.~\ref{fig1}(b) shows images of the transverse intensity distribution of the transmitted pump beam, in the near field (left) and the far-field (right). We observe for these parameters contrasted patterns with a clear hexagonal symmetry. However, here the near-field patterns are always divided into several domains with different orientations of the hexagons, and we never observe the long-range order typically associated with the optomechanical nonlinearity~\cite{Labeyrie2014}. Using parameters different from the optimal set increases the number of such domains, rendering the hexagonal symmetry less obvious (see Fig.~\ref{fig2}).

A key feature of the observed instability is the disappearance of the patterns for large pump intensity (typically $> 2$ W/cm$^2$). This behavior is illustrated in Fig.~\ref{fig2}. Just above threshold (a), the patterns appear in a restricted area around beam center. When the pump power is increased, the patterns gain in contrast and spatial extent (b). A further increase of the intensity leads to a progressive blurring of the patterns inside an area around beam center, (c) and (d). Note that heating effects due to the increase of pump intensity are negligible because of the short duration of the pump pulse. This pattern blurring is qualitatively different from what is observed e.g. for the polarization instabilities in hot vapors where no saturation is observed~\cite{Lange1998}. In our situation, this saturation is intrinsic to the 2-level description of the atom-light interaction as can be seen in Eq.~\ref{index} where $n \rightarrow 1$ as $s \rightarrow \infty$.

\begin{figure}
\begin{center}
\resizebox{1.0\columnwidth}{!}{\includegraphics{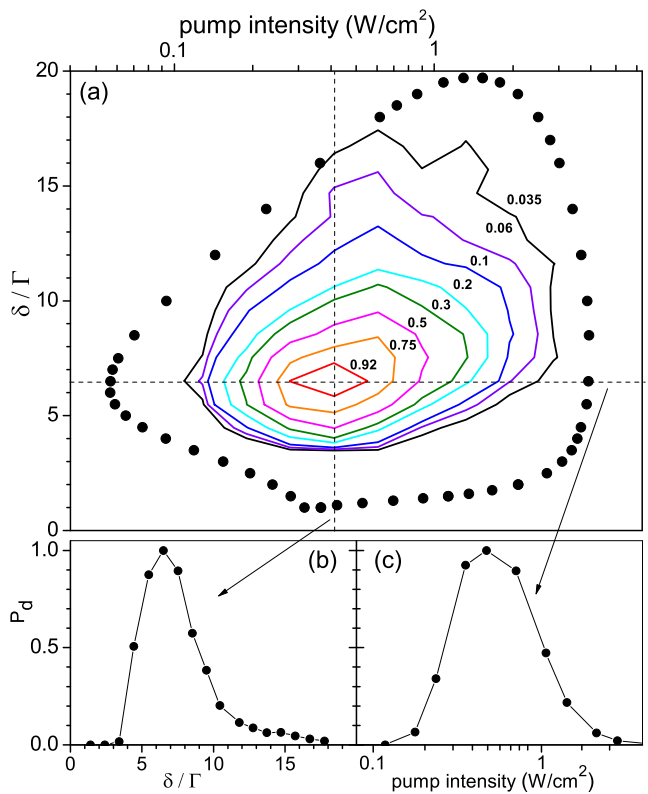}} 
\caption{(Color on line) (a) 2-level instability domain ($\delta > 0$). We measure $P_d$ (see text) as a function of $\delta$ and $I$, and plot the data as isolines. Note the logarithmic horizontal scale. The dots indicate the theoretical threshold for the instability (see text). (b) and (c) show cuts through the 2D chart of (a) as indicated by the dashed lines.}
\label{fig3}
\end{center}
\end{figure}

We have investigated the range of parameters where the instability can be observed. The result of this study is summarized in Fig.~\ref{fig3}(a), where we plot the ``diffracted power'' $P_d$ as a function of pump detuning ($\delta > 0$) and intensity. $P_d$ is obtained through the following procedure. We first record 30 successive near-field images of the patterns like that shown in Fig.~\ref{fig1}(b), to collect a representative sample of shot-to-shot pattern fluctuations. We then select an area around beam center (of diameter $w/2$) and perform a 2D numerical Fourier Transform (FT). The FT images are then summed, and we extract from the resulting averaged FT image the power in the pattern mode. This quantity is normalized to the power inside the undiffracted beam (central peak in the FT image). We then perform the same operation on images of the pump beam, obtained without atoms. This yields the background power in the pattern mode, due to the residual rugosity in the pump's intensity profile, which is subtracted from the data obtained in the presence of atoms. In addition, in Fig.~\ref{fig3} $P_d$ values are scaled such that the maximal value (obtained for $\delta = 6.5~\Gamma$ and $ I = 0.47$ W/cm$^2$) is 1.

On the blue side of the transition, we observe the patterns between roughly $\delta = 3.5~\Gamma$ and $17~\Gamma$ (Fig.~\ref{fig3}(b)). For smaller values of $\delta$, the patterns vanish quite abruptly. In this small-$\delta$ regime, the cloud is optically-thick with two important consequences: first, the strong absorption considerably reduces the magnitude of the feedback; second, there is a large amount of \emph{scattered light} with a quite homogeneous spatial distribution, which is expected to blur the transverse field modulation responsible for the instability. For large detunings the patterns also disappear but much more gradually, because of the decrease $\propto 1/\delta$ of the refractive index. We observe a well-defined lower-intensity threshold for the instability, around 0.16 W/cm$^2$ for $\delta = 6.5~\Gamma$ (Fig.~\ref{fig3}(c)). This threshold is substantially higher than observed for longer pump pulses, where the optomechanical mechanisms sets in~\cite{Labeyrie2014}. Also, we found that a minimum OD of around 100 is required to observe the 2-level patterns, while this threshold can be considerably lower for optomechanical patterns~\cite{Labeyrie2014}. The saturation of the nonlinearity results in a gradual vanishing of the instability for large pump intensity ($I > 2$ W/cm$^2$ for $\delta = 6.5~\Gamma$, which corresponds to $s \approx 7$). 

We compare in Fig.~\ref{fig3}(a) our experimental data with the theoretical instability threshold (dots) as obtained using a 2-level, thin-medium model. This model is based on the approach of~\cite{Muradyan2005}, extended to the case of the feedback mirror configuration and with the inclusion of absorption. We also included in our model the longitudinal intensity modulation due the interference of the incident and retro-reflected pumps, but without the approximation used in~\cite{Muradyan2005}. As can be seen, the qualitative and quantitative agreement is rather satisfactory. We speculate that the discrepancy at small $\delta$ may come from the scattered light (not included in the model), as discussed above.

On the red side of the transition ($\delta < 0$), we only observe poorly-contrasted structures without clear symmetry. The characteristic spatial scale of these structures is roughly twice that on the blue side. Their domain of observation in ($\delta$, $I$) space approximately mirrors that of the patterns on the blue side. A full theoretical explanation for this red-blue asymmetry is still lacking. We believe that its origin lies in nonlinear propagation effects taking place \textit{inside} the cloud. This may not come as a surprise since such effects have been observed in the past in such large cold atom clouds~\cite{Labeyrie2011}. In that work, we investigated the self-trapping of a Gaussian beam of small waist (20 $\mu$m) for $\delta > 0$, which resulted in a roughly constant transverse size of the beam as it propagated inside the cloud. It is thus reasonable to speculate that if an array of bright spots such as seen in the transverse intensity distribution of Fig.~\ref{fig1}(b) forms inside the medium, it will be stabilized by self-focusing for $\delta > 0$. On the contrary, for $\delta < 0$ self-defocusing will tend to blur these structures. For these effects to play a role, one requires the Rayleigh length corresponding to the transverse size of the bright intensity spots to be smaller than the length of the medium. This condition imposes a size for the spots of a few tens of microns, which is what we typically observe.

The Talbot effect and the associated periodic passage between phase and intensity modulation~\cite{d'Alessandro1991} is at the heart of the transverse instability discussed in this paper. Since we operate well detuned from resonance, a transverse intensity pattern mainly induces a transverse phase modulation. Propagation to the mirror and back can convert this into transverse intensity modulation of the backward field, hence phase-modulating the forward field, and so on. For a mirror distance $d$, the transverse pattern wavelength $\Lambda$ for which this is optimum obeys $\Lambda^2 = \lambda d / (N + 1/4)$ in the case of a thin medium and of a self-focusing nonlinearity~\cite{Firth1990, Ciaramella1993}. In this expression, $N$ is an integer of same sign as $d$. Furthermore, the Talbot effect implies that instability thresholds are periodic in mirror distance $d$ with period $\Lambda^2 / \lambda$.

\begin{figure}
\begin{center}
\resizebox{1.05\columnwidth}{!}{\includegraphics{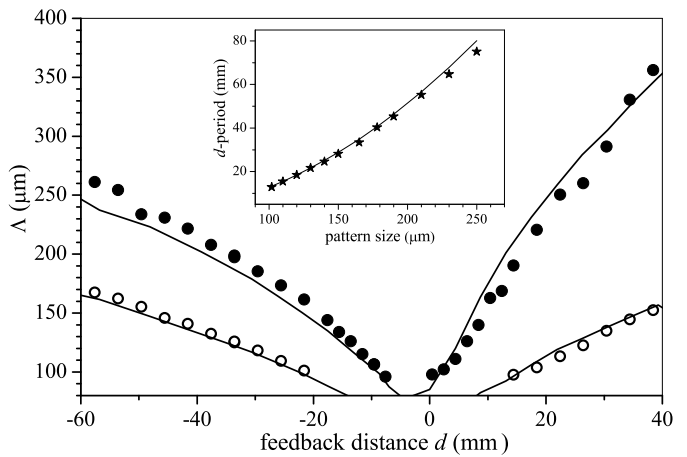}} 
\caption{Talbot effect and pattern size ($\delta > 0$). We measure $\Lambda$ (see text) as the feedback mirror distance $d$ is varied (dots: lowest $q$-mode, circles: higher $q$-mode). The line is the prediction of a thick-medium model. The insert shows the measured $d$-period as a function of the pattern size (stars), together with the Talbot effect prediction (line).}
\label{fig4}
\end{center}
\end{figure}

This tunability and $d$-periodicity of the pattern scale, features specific to the single-mirror feedback scheme, are illustrated with Fig.~\ref{fig4} where we plot $\Lambda$ (measured by far-field imaging of the transmitted pump, see Fig.~\ref{fig1}(b)) versus $d$. The dots correspond to the lowest-q mode, where $q = 2 \pi / \Lambda$. The circles correspond to the next higher-q mode, which is observed only for large $\left|d\right|$. The bold lines are predictions of a thick-medium model, similar to that of Fig.~\ref{fig3}(a) but including the propagation inside the cloud and neglecting absorption. The Talbot periodicity appears through the fact that the same $q$, and hence $\Lambda$, is observed for periodically-spaced values of $d$, the $d$-period being $\Lambda^2 / \lambda$. This is verified in the insert, where the experimentally observed $d$-period is plotted against $\Lambda$ (dots) and compared to the expression above (line). The overall agreement between experiment and theory is very satisfactory, for all instability branches, confirming the validity of the Talbot picture in our situation.

We demonstrated in this paper the existence of a pattern-forming optical instability in a cloud of cold atoms, based only on the 2-level electronic nonlinearity. In this paradigmatic situation, we were able to observe the disappearance of the instability at high optical power, due to the saturation of the nonlinearity. This work demonstrates the interest of cold atomic samples for the field of nonlinear optics and pattern formation, motivated by the fact that several nonlinear mechanisms coexist and can be selected and studied independently. Understanding and controlling these various mechanisms constitutes an important step in the future prospect of extending these experiments to degenerate quantum gases, where the simultaneous self-organization of light and matter can lead to a rich class of physical phenomena~\cite{Gopalakrishnan2009}.

\acknowledgments{The Strathclyde group is grateful for support by the Leverhulme Trust and EPSRC, the collaboration between the two groups is supported by the Royal Society (London). The Sophia Antipolis group is supported by CNRS, UNS, and R\'{e}gion PACA. We acknowledge fruitful discussions with Dan Gauthier. WJF also acknowledges sharing of unpublished work by M. Saffman.}

\end{document}